\documentclass[a4paper,10pt]{article}
\pdfoutput=1
\usepackage[utf8]{inputenc}
\usepackage{amsmath}
\usepackage{caption}
\usepackage{subcaption}
\usepackage{amssymb}
\usepackage{color}
\usepackage{cancel}
\usepackage{framed}
\usepackage{comment}
\usepackage[margin=1in]{geometry}
\usepackage{mathtools}

\newcommand{\nn}{\nonumber}

\def\a{\alpha}

 \def\G{\Gamma} 

 \newcommand{\Ncal}{{\mathcal N}}
 \newcommand{\Ocal}{{\mathcal O}}
\newcommand{\Fcal}{{\mathcal F}} \newcommand{\Dcal}{{\mathcal D}}

\newcommand*{\affaddr}[1]{#1} 
\newcommand*{\affmark}[1][*]{\textsuperscript{#1}}

\def\Cincy{Department of Physics, University of Cincinnati, Cincinnati, Ohio 45221, USA}
\def\Fermilab{Fermi National Accelerator Laboratory, P.O. Box 500, Batavia, IL 60510, USA}

\begin{document}

\date{}
\title{\Large\bfseries Collisions of Dark Matter Axion Stars with Astrophysical Sources}

%

\author{%
Joshua Eby\affmark[*]\affmark[$\dag$], 
Madelyn Leembruggen\affmark[*], 
Joseph Leeney\affmark[*], \\
Peter Suranyi\affmark[*], and 
L.C.R. Wijewardhana\affmark[*] \vspace{.3cm} \\
{\it\affaddr{\affmark[*]\Cincy}}\\
{\it\affaddr{\affmark[$\dag$]\Fermilab}}\\
}

\begin{titlepage}
\maketitle
\thispagestyle{empty}

\vskip 2 cm

\begin{abstract}
If QCD axions form a large fraction of the total mass of dark matter, then axion stars could be very abundant in galaxies. As a result, collisions with each other, and with other astrophysical bodies, can occur. We calculate the rate and analyze the consequences of three classes of collisions, those occurring between a dilute axion star and: another dilute axion star, an ordinary star, or a neutron star. In all cases we attempt to quantify the most important astrophysical uncertainties; we also pay particular attention to scenarios in which collisions lead to collapse of otherwise stable axion stars, and possible subsequent decay through number changing interactions.  Collisions between two axion stars can occur with a high total rate, but the low relative velocity required for collapse to occur leads to a very low total rate of collapses. On the other hand, collisions between an axion star and an ordinary star have a large rate, $\Gamma_\odot \sim 3000$ collisions/year/galaxy, and for sufficiently heavy axion stars, it is plausible that most or all such collisions lead to collapse. We identify in this case a parameter space which has a stable region and a region in which collision triggers collapse, which depend on the axion number ($N$) in the axion star, and a ratio of mass to radius cubed characterizing the ordinary star ($M_s/R_s^3$). Finally, we revisit the calculation of collision rates between axion stars and neutron stars, improving on previous estimates by taking cylindrical symmetry of the neutron star distribution into account. Collapse and subsequent decay through collision processes, if occurring with a significant rate, can affect dark matter phenomenology and the axion star mass distribution.
\end{abstract}

\end{titlepage}

\section{Introduction}

Axion stars \cite{Tkachev,KolbTkachev,HoganRees} are macroscopic bound states of axion particles \cite{PQ1,PQ2,Weinberg,Wilczek,DFS,Zhitnitsky,Kim,Shifman}, and their existence can have astrophysical or cosmological implications \cite{Khlopov1,Khlopov2,Khlopov3,Khlopov4,Khlopov5}. In particular, axions could form all or part of dark matter in the universe, potentially in the form of axion stars \cite{Preskill,Sikivie1,Davidson,DF,Holman,Sikivie2}. Axions could also be connected to leptonic mass hierarchy and mixings as well \cite{Ahn,Ahn2}. The masses of dilute, or weakly bound, axion stars \cite{BB,ESVW,Guth} are bounded above by considerations of gravitational stability \cite{ChavanisMR,ChavanisMR2}.

The endpoint of collapse of a weakly bound axion star whose mass exceeds the critical value $M_c$ has recently received a lot of attention. By analyzing the energy functional, the author of \cite{ChavanisCollapse} found that boson stars which possess attractive self-interactions collapse to black holes when $M$ exceeds $M_c$. Using a similar method, some of us \cite{ELSW} concluded that the full axion potential, which contains both attractive and repulsive interactions, is bounded from below, and the full energy functional is minimized at a dense radius $R_D$. We concluded that the endpoint of collapse for an axion star was a dense state, but with a radius still larger than the corresponding Schwarzschild radius $R_S$. The possibility of such dense states were also proposed earlier by \cite{BB,Braaten}, and if they exist in nature, they can have interesting phenomenological consequences \cite{HAS}.

Dense states for axion stars have large binding energies, and following our analysis of axion star decay in \cite{Lifetime}, we also suggested in \cite{ELSW} that during collapse a large number of relativistic axions are emitted in what is often called a Bosenova \cite{Bosenova}. The recent work of \cite{Tkachev2016} and \cite{Marsh2016}, using very different methods, seem to similarly indicate that relativistic axions are emitted from collapsing axion stars. The non-relativistic effective field theory of axion stars, as outlined in \cite{Braaten2}, can also have sensitivity to unique decay signatures, and such rates increase with the density of the axion star as well \cite{BraatenRel}. The dominant mechanism for the emission of relativistic axions is the subject of current debate, and we will not attempt to resolve it here. However, a recent paper on oscillon decay \cite{Oscillon} supports the mechanism suggested in \cite{Lifetime} of decay through emission of a single relativistic axion. A consensus seems to have emerged that as binding energy of an axion star increases, its decay rate through number changing interactions increases rapidly. This condition is satisfied by collapsing axion stars.

Collisions of axion stars with astrophysical sources could occur with a relatively high rate, especially if axion stars compose a large fraction of dark matter. Because collisions can change the energy functional for a dilute axion star, they can lead to unique collapse scenarios which, in turn, can suggest high rates of relativistic axion emission. With this in mind, it is interesting to analyze collisions of dilute axion stars with two potentially copious astrophysical sources: ordinary stars and other axion stars. 

Axions couple at loop-level to photons, which allows decay of free or condensed axions through a process $a\rightarrow2\gamma$, but this rate is believed to be small enough to be ignored on cosmological timescales \cite{TkachevPossibility}. However, in collisions with neutron stars, strong magnetic fields can stimulate these interactions, leading to bursts of photons that are potentially observable \cite{Iwazaki,BarrancoNS,TkachevFRB,IwazakiFRB}. The idea that such collisions could lead to the observed Fast Radio Bursts \cite{Lorimer,Keane,Thornton,Spitler}, which appeared originally several years ago, has been investigated as recently as this year \cite{Raby}. Because of the unique detection signatures arising from these collisions, we also revisit the calculation of the collision rate of axion stars with neutron stars.

In Section \ref{VarSec}, we review the variational method for determining the macroscopic parameters describing axion stars, which is used to analyze the axion star energy functional. We estimate collision rates of axion stars with other axion stars (Section \ref{CollAS}), and with ordinary stars (Section \ref{CollOS}), and in both cases, map the parameter space for collapse. In Section \ref{NSsec}, we calculate the collision rate of axion stars with neutron stars. We conclude in Section \ref{ConcSec}.

We will use natural units throughout, where $\hbar=c=1$.

\section{Variational Method for Axion Stars} \label{VarSec}
Axion self-interactions can be described by the low-energy potential \cite{ELSW,Braaten}
 \begin{equation} \label{Weq}
 W(\psi) = m^2\,f^2\Big[1 - \frac{\psi^*\,\psi}{2 m f^2}
	  - J_0\Big(\sqrt{\frac{2\psi^*\,\psi}{m f^2}}\Big) \Big],
 \end{equation}
 where $m$ and $f$ are the mass and decay constant of the axion, and $\psi$ is the low-energy wavefunction describing an $N$-particle condensate of axions. For QCD axions, typical values are $m=10^{-5}$ eV and $f=6\times10^{11}$ GeV, which implies the ratio $\delta\equiv f^2/M_P{}^2 = \mathcal{O}(10^{-14})\ll1$, which will be used in what follows.\footnote{We denote the Planck mass $M_P=G^{-1/2}=1.22\times10^{19}$ GeV, where $G$ is Newton's gravitational constant.} The total self-energy of the axion star (hereafter ``ASt'') is
 \[
  E_{AS}(\psi) = \int_0^R d^3r\,\Big[\frac{1}{2m}|\nabla \psi|^2
		+ \frac{1}{2}V_{grav}|\psi|^2 
	+ W(\psi) \Big],
 \]
 where $R$ is the radius of the star. The gravitational potential is taken to be Poissonian, and thus satisfies
 \[
  \nabla^2\,V_{grav} = 4\,\pi\,G\,m^2\,|\psi|^2.
 \]
 
 The self-interaction potential in eq. (\ref{Weq}) is typically expanded in powers of the axion wavefunction $\psi$. Following \cite{ELSW}, we expand $W(\psi)$ and truncate at the next-to-leading order, including both the attractive $(\psi^*\psi)^2$ and a repulsive $(\psi^*\psi)^3$ interaction. Because we will later consider possible ASt collapse, we believe that both of these interaction terms are of crucial importance.
 
We will use a variational ansatz of the form
\begin{equation} \label{ansatz}
    \psi(r) = 
  \begin{dcases}
    \sqrt{\frac{4\,\pi\,N}{(2\pi^2-15)R^3}}\cos^2\Big(\frac{\pi\,r}{2\,R}\Big),& 		r\leq R\\
    0,              & r>R
  \end{dcases}
\end{equation}
which has a well-defined edge at $r=R$, to describe the ASt wavefunction. The radius $R$ and particle number $N$ of the ASt can be rescaled as \cite{ELSW}
\begin{equation} \label{Scaling}
 \rho = \frac{m\,f}{M_P}\,R, \qquad n = \frac{m^2}{M_P\,f}\,N,
\end{equation}
giving a total self-energy of
\begin{equation} \label{Energyeq}
 E_{AS}(\rho) = m\,N\,\delta\Big[\frac{A}{\rho^2}
	-\frac{B\,n}{\rho}
	-\frac{C\,n}{\rho^3}
	+\frac{D\,\delta\,n^2}{\rho^6}\Big]
\end{equation}
 where 
 \begin{align}
  A &= \frac{\pi^2(2\pi^2-3)}{6(2\pi^2-15)} \approx 5.8
  &B& = \frac{(3456\pi^4-13800\pi^2-166705)}{1440(2\pi^2-15)^2} \approx 1.0 
	\nn \\
  C &= \frac{35\pi(24\pi^2-205)}{2304(2\pi^2-15)^2} \approx .068
  &D& = \frac{77\pi^2(600\pi^2-5369)}{691200(2\pi^2-15)^3} \approx .0057 \nn
 \end{align}  
 for the cosine ansatz of eq. (\ref{ansatz}).\footnote{Note that the structure of eq. (\ref{Energyeq}) is ansatz-independent, but the values of the coefficients $A,B,C,D$ are not.} The terms in eq. (\ref{Energyeq}) correspond to kinetic, self-gravitational, attractive interaction, and repulsive interaction energies, respectively. Higher-order interaction terms neglected here are proportional to $(\delta/\rho^3)^k$ with $k>2$, i.e. suppressed by high powers of $\delta\ll 1$.
 
\begin{figure}[t]
  \begin{center}
   \includegraphics[scale = 1]{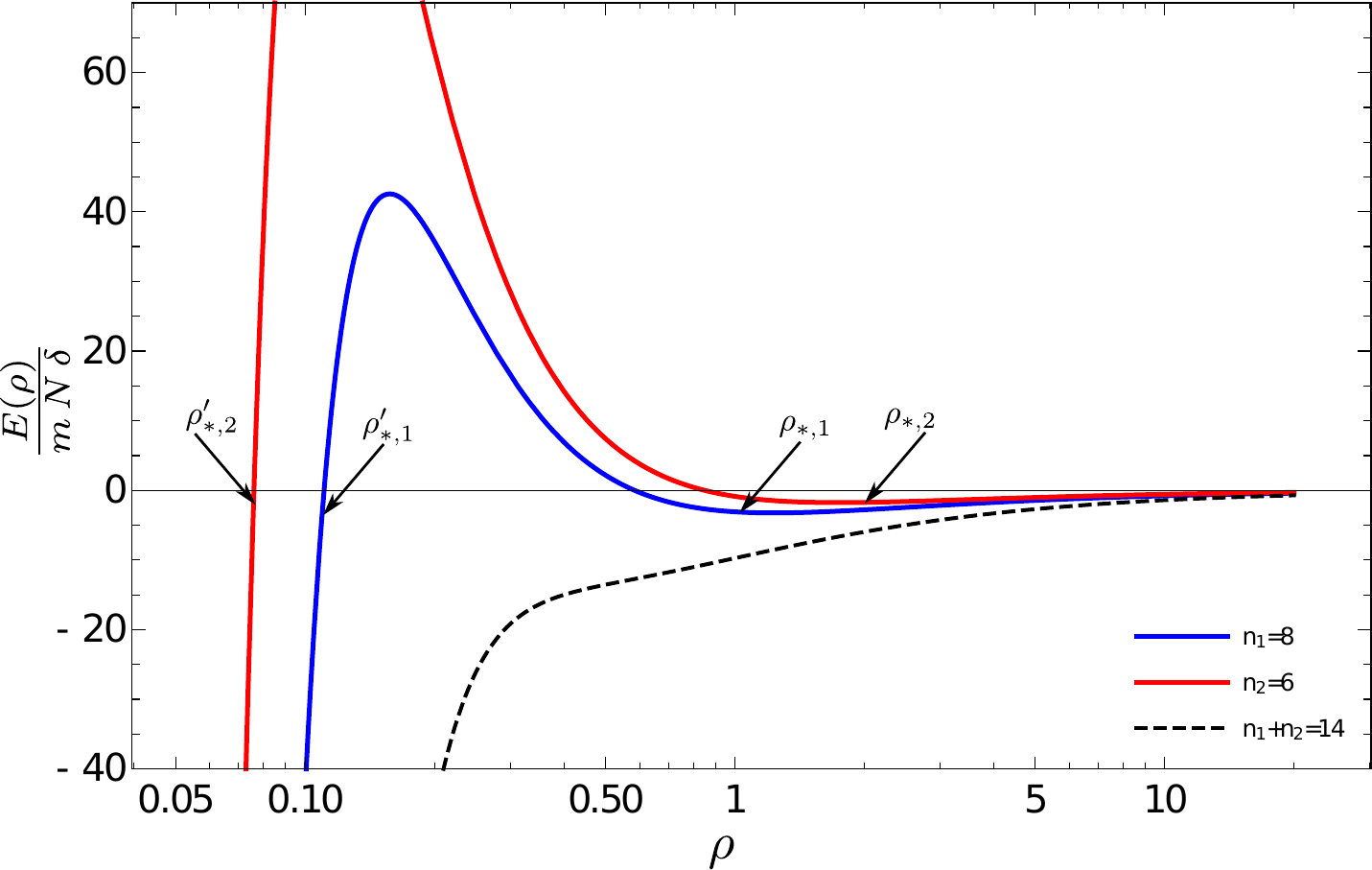}
  \end{center}
   \caption{The two solid lines denote the rescaled energy functionals of two stable ASts in the vicinity of $\rho_*$, which have $n_1=8$ (blue) and $n_2=6$ (red). The points labeled in the graph are: the local minimum of ASt 1 (2), $\rho_{*,1}$ ($\rho_{*,2}$), and the isoenergetic point on the left of the maximum of ASt 1 (2), $\rho_{*,1}'$ ($\rho_{*,2}'$). The black, dashed curve is the full energy of these two ASts occupying the same volume, which has no minimum in this range of $\rho$, because the effective particle number $n_{eff} = n_1+n_2 > n_c$; thus, during such a collision, the ASts begin to collapse.}
   \label{EnInStarFig}
  \end{figure}

 Near $\rho = \mathcal{O}(1)$, there exists a local minimium of the energy as long as $N<N_c$, a critical value of the particle number. In this range, the energy is well-approximated by only the first three terms in eq. (\ref{Energyeq}), and has a local minimum at a rescaled radius
 \begin{equation} \label{rhostar}
  \rho_* = \frac{A}{B\,n}\Big[1 - \sqrt{1 - \frac{3\,B\,C}{A^2}n^2}\Big].
 \end{equation}
 This local minimum is illustrated in Figure \ref{EnInStarFig} for two different values of $n$. One can easily read off the critical rescaled particle number $n_c = A/\sqrt{3\,B\,C}$; the energy of an ASt with effective particle number $n>n_c$ is also illustrated in Figure \ref{EnInStarFig}. Such states are referred to in the literature as \emph{dilute} ASts, and correspond to small binding energies. This is parameterized by a parameter $\Delta = \sqrt{1 - \varepsilon^2/m^2}$, where $\varepsilon$ is the energy per axion. For dilute ASts, $\Delta = \Ocal(\sqrt{\delta})$, which is $\Ocal(10^{-7})$ for the QCD axion parameters we use in our analysis \cite{ESVW}.
 
 At any rescaled particle number $n$, the global minimum of the energy in eq. (\ref{Energyeq}) lies at a very small radius $\rho_D \ll 1$ \cite{ELSW}. At these small radii, the last two terms of eq. (\ref{Energyeq}) dominate and the energy minimum lies approximately at a rescaled radius of
 \begin{equation} \label{rhoD}
  \rho_D \approx \Big(\frac{2\,D\,n\,\delta}{C}\Big)^{1/3}.
 \end{equation}
 For QCD axions with the cosine ansatz, $R_* \approx 200$ km and $R_D \approx 7$ meters when $n\approx n_c\approx 12.6$. Note $n=n_c$ corresponds to the maximum mass of a dilute ASt, which is $M_c \equiv m\,N_c \approx 10^{19}$ kg. An ASt with $n>n_c$ (as illustrated by the dashed line in Figure \ref{EnInStarFig}) possesses only a dense energy minimum at $\rho_D\ll 1$. Such a state is referred to as a \emph{dense} ASt, and has large binding energy $\Delta = \Ocal(1)$.
 
 There are various mechanisms for stimulating collapse from $R_*$ to $R_D$. For example, if ASts form against a dilute background of free axions, they could accrete such axions, thereby acquiring masses $M>M_c$ and triggering collapse; however, it is not clear how efficiently such accretion would occur. We show below that collapse could also be catalyzed by interactions between ASts and other astrophysical sources, including stellar matter or other ASts. ASts can also be converted efficiently to photons through interactions with neutron stars and their strong magnetic fields 
 \cite {Iwazaki,BarrancoNS,TkachevFRB,IwazakiFRB,Raby}. We are thus motivated to investigate the rate of collisions of ASts with these sources, how a collision could alter the energy landscape, and how the population of ASts might change as a result.
 
 ASts collapsing from the dilute radius $R_*$ to the dense one $R_D$ move quickly from small to large binding energies \cite{ELSW}. When the binding energy becomes large, the rate of number-changing interactions in the ASt grows quickly, and so the ASt emits many relativistic axions as it collapses \cite{Lifetime,Tkachev2016,Marsh2016}. In the end, we will speculate about the observable effects of ASt decay, which could be stimulated by collisions and subsequent collapse.

\section{Collisions Between Two ASts} \label{CollAS}
\subsection{Modified Energy Functional}
 We consider first collisions between two weakly-bound ASts in a single dark matter halo. Such collisions could lead to mergers of these ASts, which can trigger collapse, because the effective energy functional is modified from eq. (\ref{Energyeq}) to 
  \begin{equation} \label{E2AS}
  E_{2AS}(\rho) = m\,(N_1+N_2)\,\delta\Big[\frac{A}{\rho^2}
	-\frac{B\,(n_1+n_2)}{\rho}
	-\frac{C\,(n_1+n_2)}{\rho^3}
	+\frac{D\,\delta\,(n_1+n_2)^2}{\rho^6}\Big],
 \end{equation}
 where $N_1$ and $N_2$ are the number of axions in each of the two stars. If the sum $n_1+n_2 > n_c$, then both stars begin to collapse, as their combined energy no longer has a local minimum at $\rho_*$.
 
 However, there is no guarantee that colliding ASts will merge; because they move with some relative velocity $v_{rel}$, they may occupy the same volume for only a finite time. Another way to say this is that, in light of the weak self-interactions of axions, it is possible that such objects would pass right through one another. Indeed, no mechanism is known for dissipating energy during the collision--with the exception of gravitational waves, but the corresponding rate of energy dissipation is negligibly small.\footnote{This is the case because the masses of QCD ASts are too small to have a significant gravitational wave output, but the situation could be different in some more generic axion theory which allows very heavy bound states. We plan to return to this point in a future work.} An important, related note is that we also ignore differences in phase for the colliding ASts, which could increase the merger rate for condensates close to being in-phase, or lead to inelastic ``bounces'' when the condensates are out of phase \cite{Cotner}. For sufficiently large velocities, these effects are likely to be negligible, but could be relevant for ASts with low relative velocities. We hope to return to this point in the near future.
 
 Thus, for the purposes of this work, we will assume that colliding ASts do not dissipate energy and become bound, and so the energy functional changes for only a finite time; we approximate this time by $t_{in} = 2(R_1+R_2)/v_{rel}$, where $R_{1}$ and $R_2$ are radii of the two ASts. Nonetheless, collapse will begin when the stars occupy the same volume, as the energy functional changes. If the ASts overlap for a sufficiently long time, then when the stars separate, they will already be gravitationally unstable and will continue to collapse.

\subsection{Collision Rates} \label{CollRatesASSec}
 In this section, we begin to examine the approximate rates for two types of astrophysical bodies colliding with each other in a single galaxy, where at least one of these objects is an ASt. We will perform this calculation in two ways: we begin by making the assumption that ASts are distributed with some constant density in galactic halos; later, we will take into account the nontrivial number density. 
 
 For two populations of astrophysical objects, the first being ASts and the other labeled by $i$, the general expression for the collision rate is
 \begin{equation} \label{GammaGeneric}
  \Gamma_i = \frac{1}{S}\int\,d^3r\,n_{AS}(\vec{r})\,n_{i}(\vec{r})
	  \langle\sigma\,v\rangle_i
 \end{equation}
 where $n(\vec{r})$ denotes the number density of some population of astrophysical objects, $\sigma_i$ is the cross section for a collision, and $v$ is the relative velocity between the objects. The ``symmetry factor'' $S=2$ if the populations are the same type, and $S=1$ otherwise. Generically the expectation value
 \[
  \langle\sigma\,v\rangle_i \equiv \int_0^{2\,v_{vir}}f_{MB}(v)\,\sigma_i
	  \,v\,d^3v
 \]
 is an average over velocities in the halo, using a Maxwell-Boltzmann distribution of velocities
 \begin{equation} \label{fMB}
  f_{MB}(v) = f_0\,\exp\Big(-\frac{v^2}{v_{vir}^2}\Big).
 \end{equation}
 The distribution is normalized so that $\int_0^{2v_{vir}}f_{MB}(v)\,d^3v = 1$, and $v_{vir}$ is the virial velocity in the halo.
 
 In the special case where both distributions are trivial, that is, that the objects are distributed with constant density throughout the halo, eq. (\ref{GammaGeneric}) simplifies greatly. In that case, $n_i(\vec{r}) = N_i/V_{gal}$ is constant, where $N_i$ is the total number of objects $i$ present and $V_{gal}=4\pi R_{gal}^3/3$ is the volume of the galaxy. Then the integration is trivial, and the rate for collisions of a species $i$ with ASts in a dark matter halo is
\begin{equation} \label{rate}
 \hat{\Gamma}_i = \frac{N_{AS}\,N_i\,\sigma_i\,v}{\frac{4\pi}{3}R_{gal}^3}.
\end{equation}
 This expression is significantly simpler than eq. (\ref{GammaGeneric}), and often gives a good order of magnitude estimate, but as we will see in later sections, it sometimes vastly underestimates collision rates.
 
 In this work we will use data for the Milky Way to approximate the collision rates for ``typical'' galaxies. For this reason we set $R_{gal}\approx 10^{5}$ lightyears, which is the radius of the Milky Way.  For the cross section, we will use 
 \[
  \sigma_i = \pi\,(R_i+R_{AS})^2\,\Big(1 
	  + \frac{2\,G\,(M_i+M_{AS})}{v^2\,(R_i+R_{AS})}\Big),
 \]
 with $R_i$ the radius of the body which collides with the ASt, and $M_i$ is its mass. This is the geometric cross section modified by an enhancement factor to account for classical capture through gravitational effects. 
 
 Note that we will assume in what follows that the initial configuration of the colliding ASt is the dilute state (not the dense state). This is because, first, the collision rates for dense ASts can be suppressed in some cases by the very small factor $(R_*/R_D)^2\sim 10^{-10}$, as defined by eqs. (\ref{rhostar}) and (\ref{rhoD}), compared to dilute ASts; and second, because otherwise stable dilute ASts can collapse as a result of collisions, making them additionally interesting.
 
 The number of ASts in the galaxy $N_{AS}$ can be estimated, albeit roughly, by assuming that the total mass $M_{DM}\approx10^{12}M_\odot$ of the dark matter in the Milky Way consists of ASts with a fixed mass of $M_{AS}=M_c \approx 10^{19}$ kg. ($M_\odot=2\times10^{30}$ kg is a solar mass.) In truth, there are two prominent effects that would modify this estimate. First, dark matter may consist only partially of ASts, the rest potentially being in a dilute background of axions or some other dark particles; we represent this effect by a multiplicative factor $0\lesssim\Fcal_{DM}\lesssim1$, so that the total dark matter mass in ASts is $\Fcal_{DM}M_{DM}$. Secondly, ASts likely have a spread in their mass distribution, and so some ASts will have masses smaller than the maximum $M_c$; we introduce a second factor $0\lesssim\Fcal_{AS}\lesssim1$, so that the average mass of a single ASt is $\Fcal_{AS}M_{AS}$. These competing effects can, of course, compensate each other in the calculation of $N_{AS}$. The total number of ASts in the Milky Way is
 \[
  N_{AS} = \frac{\Fcal_{DM}M_{DM}}{\Fcal_{AS}M_{AS}} 
	  \approx \frac{\Fcal_{DM} 10^{12}M_\odot}{\Fcal_{AS} 10^{19} \text{ kg}}
	  = 2\times10^{23} \frac{\Fcal_{DM}}{\Fcal_{AS}}.
 \]

 For the specific case of collisions of two ASts, we use the radius $R_{AS} = 200$ km and the constant velocity $\bar{v}=300$ km/s; the enhancement factor
 \[
  \frac{2\,G\,M_{AS}}{\bar{v}^2\,R_{AS}} \approx 10^{-7}
 \]
 is negligible in the ASt cross section in this case. Then a constant-density estimate of the rate in eq. (\ref{rate}) is
 \begin{equation} \label{CD_Rate_AS}
  \hat{\Gamma}_{AS} \approx 3\times10^{7} 
	\Big(\frac{\Fcal_{DM}}{\Fcal_{AS}}\Big)^2 
	      \frac{\text{collisions}}{\text{year}\cdot\text{galaxy}}.
 \end{equation}
 
 \begin{figure}[t]
  \begin{center}
   \includegraphics[scale=1]{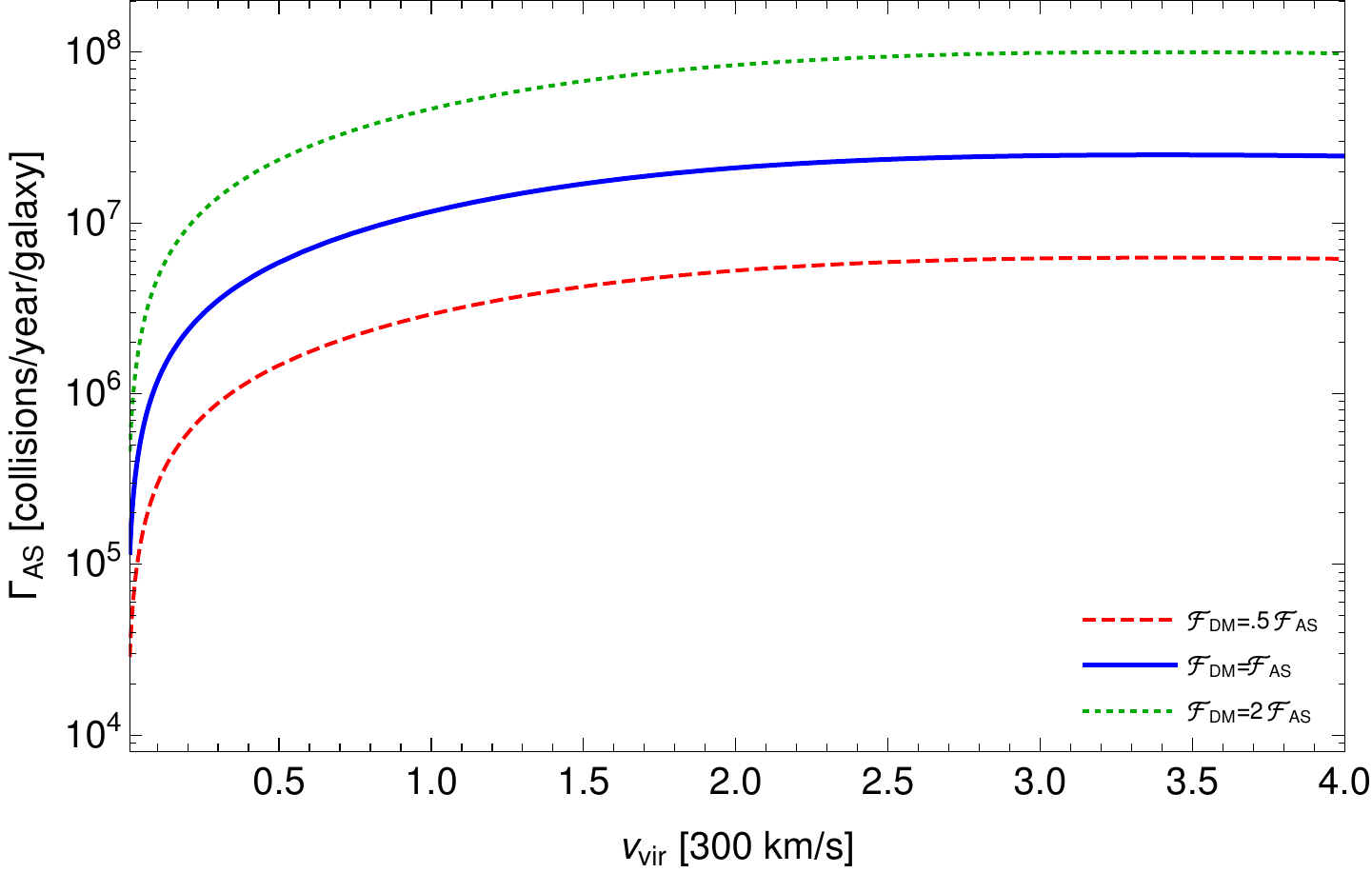}
  \end{center}
   \caption{The collision rate of two ASts as a function of $\eta=v_{vir}/(300 \text{ km/s})$, evaluated at different values of $\Fcal_{DM}/\Fcal_{AS}$.}
   \label{GammaASFig}
  \end{figure}
 
 We can improve on this rough estimate by taking into account the number density distributions of ASts. If ASts are a component of dark matter, then it is reasonable to assume that they are distributed according to the Navarro-Frenk-White (NFW) profile \cite{NFW}
 \begin{equation} \label{NFW}
  \rho_{NFW}(r) =  \frac{\rho_0}{\frac{r}{R_0}\Big(1 + \frac{r}{R_0}\Big)^2},
 \end{equation}
 where $\rho_0$ and $R_0$ are scale mass density and radius for the halo. An analysis of this kind was recently performed by \cite{Cholis}, who considered primordial black holes as dark matter. Taking the Milky Way as a sample galaxy, we use the values of $\rho_0$, $R_0$, and virial mass $M_{vir}$ from \cite{NFWconst}. These values are reported in Table \ref{MWtab}. We will also take into account velocity dispersion by assuming the Maxwell-Boltzmann distribution of eq. (\ref{fMB}), though the precise value of $v_{vir}$ is somewhat uncertain. We introduce an $\Ocal(1)$ factor $\eta$ representing the deviation of $v_{vir}$ from the value $\bar{v}=300$ km/s used above. That is,
 \[
  v_{vir} = \eta \, \bar{v},
 \]
 so fixing the value of $M_{vir}$ from \cite{NFWconst}, we have a virial radius of
 \[
  R_{vir} = \frac{G\,M_{vir}}{\eta^2\bar{v}^2} 
	\approx \frac{2.22\times10^{18} \text{ km}}{\eta^2}.
 \] 
 
  The velocity-averaged cross section is   
  \[
  \langle \sigma\,v\rangle_{AS} = \int_0^{2v_{vir}}f_{MB}(v)\,
		    \sigma_{AS}\,v\,d^3v
	  \approx 1.6\times10^8\,\eta \frac{\text{km}^3}{\text{sec}},
 \]
 implying a total rate for collisions 
 \begin{align} \label{NFWRate}
  \Gamma_{AS} &= \frac{1}{2} \int_0^{R_{vir}} 4\pi r^2 
	  \Big(\frac{\Fcal_{DM}\rho_{NFW}(r)}{\Fcal_{AS}M_{AS}}\Big)^2
	      \langle\sigma\,v\rangle_{AS}
	      \, dr  \nn \\
	&= \frac{2\pi\langle\sigma\,v\rangle_{AS}}{M_{AS}^2}
	\Big(\frac{\Fcal_{DM}}{\Fcal_{AS}}\Big)^2
		\int_0^{R_{vir}}r^2\rho_{NFW}(r)^2\,dr \nn \\
	&\approx 10^7 \,\eta 
	      \Big(\frac{\Fcal_{DM}}{\Fcal_{AS}}\Big)^2 
	      \Big[1 - \Big(\frac{\eta^2}{4.5+\eta^2}\Big)^3\Big]
	      \frac{\text{collisions}}{\text{year}\cdot\text{galaxy}}.
 \end{align}
 The $\eta$-dependence in the last bracket comes from the uncertainty in $R_{vir}$ in the upper bound of the integral. If $\eta=1$, we have
 \begin{equation} \label{eta=1}
  \Gamma_{AS}|_{\eta=1} = 10^7
	    \Big(\frac{\Fcal_{DM}}{\Fcal_{AS}}\Big)^2 
	    \frac{\text{collisions}}{\text{year}\cdot\text{galaxy}}.
 \end{equation}
 Thus in this case, the constant density approximation in eq. (\ref{rate}) gives a reasonable estimate, that of eq. (\ref{CD_Rate_AS}). The full decay rate $\Gamma_{AS}$ over a range of $\eta$, and for different values of $\Fcal_{DM}/\Fcal_{AS}$, is represented in Figure \ref{GammaASFig}. The dependence on $\Fcal_{DM}$ and $\Fcal_{AS}$ implies that the true collision rate could easily be larger or smaller by a few orders of magnitude from the approximate value $10^7$ collisions/year/galaxy.
 
  \begin{table}[ht]
 \begin{center}
 \begin{tabular}{| c | c |} 
  \hline
  $\rho_0$ [$M_\odot$/kpc$^3$] & $R_0$ [kpc] \\
  \hline
  $1.4\times10^{7}$ & $16$ \\
  \hline
  $M_{vir}$ [$M_\odot$] & -- \\
  \hline
  $1.5\times10^{12}$ & -- \\
  \hline
 \end{tabular}
  \begin{tabular}{| c | c | c | c |} 
  \hline
  $L_1$ [pc] & $L_2$ [pc] & $L_\odot$ [kpc] & $f$ \\
  \hline
  $2600$ & $3600$ &  $8$ & $.12$ \\
  \hline
  \hline
  $H_1$ [pc] & $H_2$ [pc] & $Z_\odot$ [pc] &  -- \\
  \hline
  $300$ & $900$ & $25$  & -- \\
  \hline
 \end{tabular}
 \\
   \begin{tabular}{| c | c | c | c | c | c | c |} 
  \hline
  $A_{0,r}$ [kpc$^{-1}$] & $\alpha$ & $A_{0,z}$ [kpc$^{-1}$] & $A_{2,z}$ [kpc$^{-1}$] & $k_1$ & $k_2^<$ & $k_2^>$ \\
  \hline
  $5\times10^{-3}$ & $1.83$ & $1.8\times10^{-5}$ & $35.6\times10^{-3}$ & $13\times10^{-3}$ & $18.4\times10^{-3}$  & $.05$ \\
  \hline
  \hline
  $A$ & $\lambda$ [kpc] & $A_{1,z}$ [kpc$^{-1}$] & -- & $b_1$ [kpc] & $b_2^<$ [kpc] & $b_2^>$ [kpc] \\
  \hline
  $95.6\times10^{-3}$ & $4.48$ & $1.87$ & -- & $12.8\times10^{-3}$ & $.03$ & $.65$ \\
  \hline
 \end{tabular}
 \caption{Astrophysical parameters describing the Milky Way, which we use in the calculation of collision rates. The table on the top-left describes the dark matter halo using an NFW profile using data from \cite{NFWconst}; the top-right table describes the distribution of stars using \cite{MilkyWay}; and the table along the bottom gives parameters describing the distribution of neutron stars using \cite{NSdist}.}
 \label{MWtab}
 \end{center}
 \end{table}
 
\subsection{Collapse Time}
The total energy of two ASts as they pass through each other can be approximated by eq. (\ref{E2AS}), which is equivalent to eq. (\ref{Energyeq}) with $N\rightarrow N_1+N_2$. This is equivalent to an increase in the effective particle number in the total energy, while the stars occupy the same volume. If two dilute ASts have $N_1+N_2>N_c$, their combined energy functional in eq. (\ref{E2AS}) will no longer possess a dilute energy minimum at $\rho_*$, triggering the collapse process for both stars. We illustrate the change in the energy during the collision for a set of sample parameters in Figure \ref{EnInStarFig}. 

During the time the two ASts occupy the same volume, if one of them collapses sufficiently far, it will remain gravitationally unstable even after it passes through the other. A sufficient condition for this would be that its radius decreases from $\rho_*$ to $\rho_*'$, identified as the point isoenergetic with $\rho_*$ on the left branch of the energy curve (illustrated in Figure \ref{EnInStarFig}). The collapse time from $\rho_*$ to some other point $\rho_{end}$ is \cite{ELSW}
 \begin{equation} \label{tColl}
  t = \sqrt{\frac{q\,M}{2}} 
	\int_{R_{end}}^{R_*}\frac{dR}{\sqrt{E(R_0) - E(R)}},
 \end{equation}
 where 
 \[
  q = \frac{3(315 - 50\pi^2 + 2 \pi^4)}{5\pi^2(2\pi^2 - 15)} \approx .2
 \]
 for the cosine ansatz. Thus we will analyze the time $t_*$ to collapse from $\rho_*$ to $\rho_*'$ (i.e. using $R_{end} = R_*'$).
 
 Full collapse from the dilute radius $\rho_*$ to the dense radius $\rho_D$ was shown in \cite{ELSW} to last a time on the order of tens of minutes, or even a few hours, depending on the value of $N/N_c$. We find that collapse from $\rho_*$ to $\rho_*'$, catalyzed by collision of two nearly-critical ASts, takes nearly as long as a collapse all the way to $\rho_D$. This is easy to understand if one notes that the potential is shallow at the beginning of the collapse, near $\rho=\rho_*$, and becomes steeper towards the end (see Figure \ref{EnInStarFig}). Because collapses occur slowly, we conclude that colliding ASts only collapse if they move with a sufficiently low relative velocity, so that the two ASts occupy the same volume for a long enough time interval. We investigate next how small $v_{rel}$ must be for collapse to be induced.
 
 \begin{figure}[t]
  \begin{center}
   \includegraphics[scale=.75]{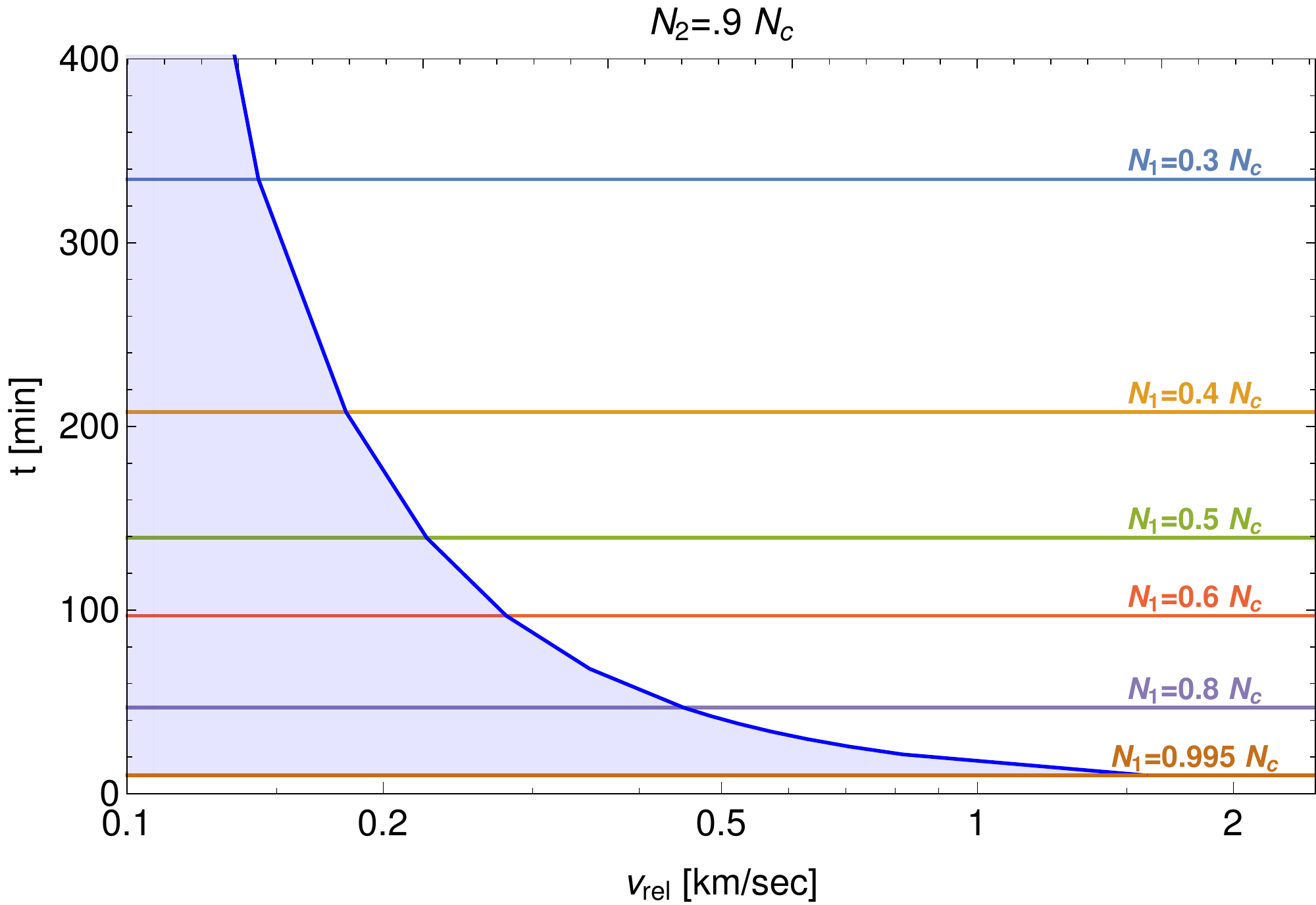}
  \end{center}
   \caption{The relevant timescales for the collision and possible collapse of two ASts. The horizontal lines are the required times $t_*$ for an ASt with $N_1$ particles to collapse from $\rho_*$ to $\rho_*'$, given that it collides with a second ASt which has $N_2$ partices at a relative velocity $v_{rel}$. The blue curve represents the approximate time $t_{in}$ that one ASt spends traversing the other. If $t_{in}<t_*$ (the blue shaded region), then the ASt collapses.}
   \label{ASAS_time}
  \end{figure}
 
 In Figure \ref{ASAS_time}, we compare the approximate collision time $t_{in}= 2(R_1+R_2)/v_{rel}$ to the time $t_*$ required for a star to collapse from $\rho_*$ to $\rho_*'$, defined in eq. (\ref{tColl}), at different values of $n$. We observe that $v_{rel}\lesssim 1$ km is required for nearly any two ASts to collapse as a result of the collision process. For the distribution of velocities in eq. (\ref{fMB}), the probability that two ASts have a relative velocity $v_{rel} \lesssim V$ is
 \[
  P(v_{rel}\lesssim V) = \sqrt{\frac{2}{9\pi}}
	    \Big(\frac{V}{v_{vir}}\Big)^3 + \Ocal\Big(\frac{V}{v_{vir}}\Big)^5
 \]
 For $V = 1$ km/sec, this gives $P(v_{rel} < 1$ km/sec$) \approx 10^{-8}$. This implies that while the collision rate in eq. (\ref{eta=1}) is high, even on the assumption that all ASts are very close to critical ($N\sim .9\,N_c$), the total rate of collapses induced by collisions of two ASts is very small,
 \begin{equation}
  \G_{collapse} \sim \G_{AS}\times P(v_{rel} < 1 \frac{\text{km}}{\text{sec}}) 
	      \approx .1 \,\Big(\frac{\Fcal_{DM}}{\Fcal_{AS}}\Big)^2 
	    \frac{\text{collapses}}{\text{year}\cdot\text{galaxy}}.
 \end{equation}
 The rate of induced collapses is even lower if the colliding ASts are lighter. We conclude that collapses induced by collisions between two ASts very likely have a negligible effect on the overal mass distribution of ASts, or more generally, of dark matter.
 
 We also note in passing that if ASts collide but do not collapse fully from $\rho_*$ to $\rho_*'$ (e.g. a scenario outside of the shaded blue region of Figure \ref{ASAS_time}), they will oscillate around their respective dilute minima after emerging from the collision. With no mechanism for damping, such oscillations would continue indefinitely.
 
  \section{Collision of an ASt with an Ordinary Star} \label{CollOS}
\subsection{Gravitational Potential Inside Star}
ASts can also collide with ordinary stars. As an ASt passes through an ordinary star, there will be an additional contribution to its gravitational energy due to the stellar potential $V_s(r)$. To estimate this contribution, consider first an ASt at rest at a point concentric with that of a star. The energy functional describing the ASt, given by eq. (\ref{Energyeq}), acquires an additional contribution from the gravitational effects of the stellar mass,
\begin{equation} \label{EGS}
 E_{GS}(R) = m\,\int \,V_s(r)|\psi(r)|^2\,d^3r.
\end{equation}
Note that we have assumed that the external gravitational source has a radius $R_s$ larger than the radius $R$ of the ASt; this is appropriate for the types of QCD axions we discuss here, but could change in some more generic axion theory which produces very large ASts. Assuming a constant density for the star and requiring continuity of the full gravitational potential across each boundary gives the unique result
\begin{equation} \label{Vstar}
    V_s(r) =
  \begin{dcases}
    \frac{G\,M_s}{2\,R_s^3}r^2 - \frac{3}{2}\frac{G\,M_s}{R_s},
	      & r\leq R_s\\
    -\frac{G\,M_s}{r},              & r>R_s
  \end{dcases}
\end{equation}
 where $M_s$ is the mass of the star.

  \begin{figure}[t]
  \begin{center}
   \includegraphics[scale=.75]{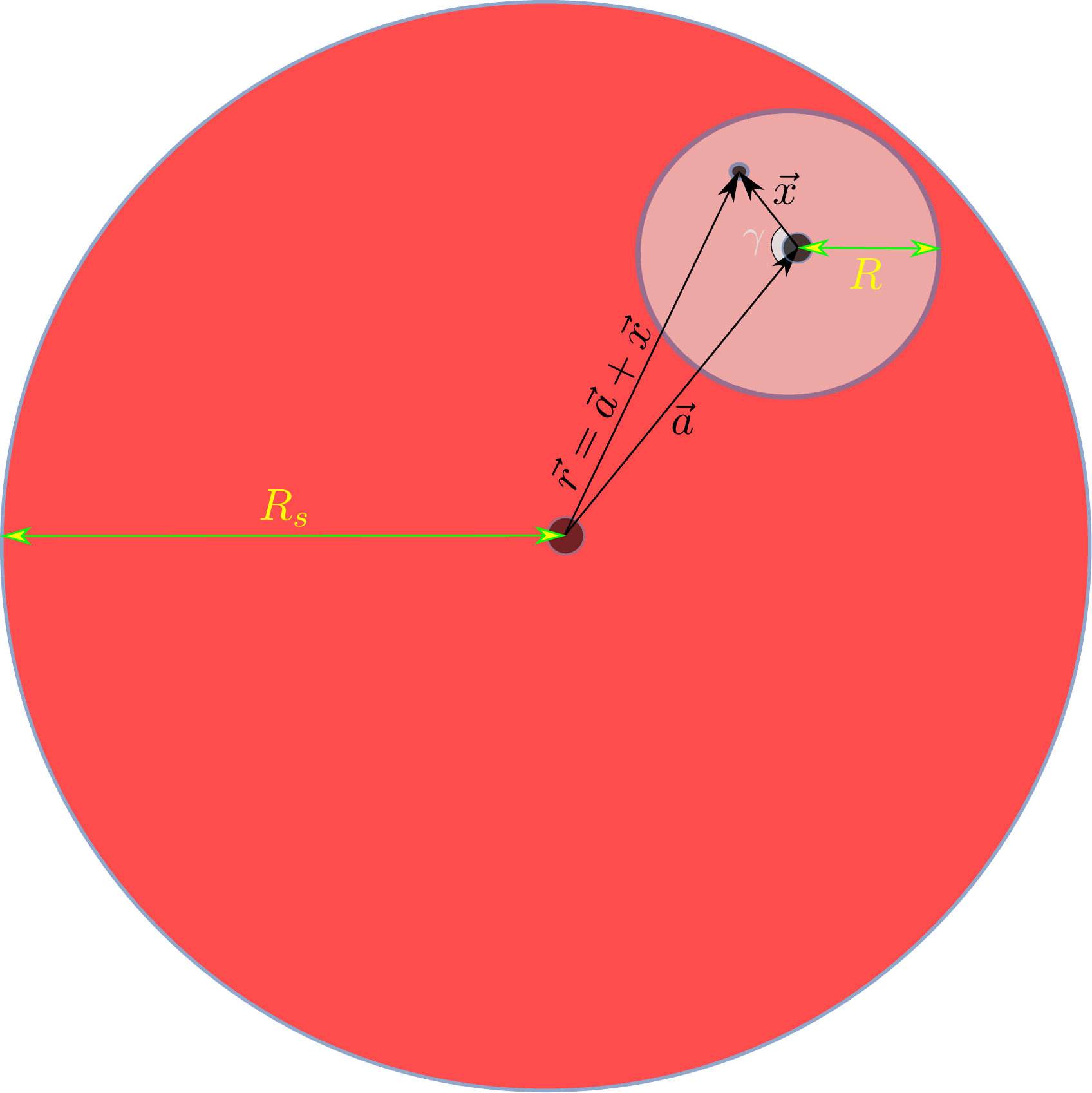}
  \end{center}
   \caption{An illustration of an ASt passing through an ordinary star. The relative radii of the ASt ($R$) and the star ($R_s$) are not to scale.}
   \label{AS_Star_Fig}
  \end{figure}

In reality, ASts and ordinary stars will collide with some relative velocity $v_{rel}$ and the gravitational potential will be changing with time. However, the change in the total energy resulting from this effect is a function of the distance from the ASt to the center of the ordinary star (not of the variational parameter $R$), and so will not affect the existence or position of an energy minimum. To see that this is so, suppose the ASt sits at a point $\vec{a}$ away from the center of the star, and let $\vec{x}$ represent a position measured with respect to the center of the ASt; then $\vec{r} = \vec{a} + \vec{x}$, as in Figure \ref{AS_Star_Fig}. Thus eqs. (\ref{EGS}) and (\ref{Vstar}) give
\begin{align} \label{EGScalc}
 E_{GS}(R) &= \frac{G\,m\,M_s}{2\,R_s^3}\int r^2 |\psi(r)|^2 d^3r
	- \frac{3}{2}\frac{G\,m\,M_s}{R_s}\int |\psi(r)|^2 d^3r \nn \\
      &= \frac{G\,m\,M_s}{2\,R_s^3}\int 
	      (\vec{a}+\vec{x})^2 |\psi(\vec a+\vec x)|^2 d^3x 
		- \frac{3}{2}\frac{G\,m\,M_s}{R_s}\,N \nn \\
      &= \frac{G\,m\,M_s}{2\,R_s^3}\int 
	      (a^2+x^2+2\vec{a}\cdot\vec{x}\cos\gamma) |\psi(\vec a+\vec x)|^2 d^3x 
		- \frac{3}{2}\frac{G\,m\,M_s}{R_s}\,N \nn \\
      &= \frac{G\,m\,M_s}{2\,R_s^3}\int_0^R 
	      (x^2+2\vec{a}\cdot\vec{x}\cos\gamma) |\psi(\vec a+\vec x)|^2 
		    x^2\,dx\,d(\cos\gamma)d\phi
		+ \frac{G\,m\,M_s}{2\,R_s^3}\,a^2\,N
		- \frac{3}{2}\frac{G\,m\,M_s}{R_s}\,N,
\end{align}
 where in the third line we introduced the angle $\gamma$ between the vectors $\vec{a}$ and $\vec{x}$. For a spherically symmetric ASt wavefunction, $\psi(\vec{x})=\psi(x)$ does not depend on $\gamma$, and so the second integral vanishes. The last two terms of eq. (\ref{EGScalc}) are constants with respect to $R$. For notational simplicity, we denote them by
 \begin{align}
  C_1 &\equiv \frac{G\,m\,M_s}{2\,R_s^3}\,a^2\,N
		- \frac{3}{2}\frac{G\,m\,M_s}{R_s}\,N \nn \\
      &= m\,N\,\delta\Big[\frac{\mu_s\,\alpha^2}{2\,\rho_s^3}
	      -\frac{3\,\mu_s}{2\,\rho_s} \Big] \nn \\
      &= m\,N\,\delta\,c_1,
 \end{align}
 where we defined the dimensionless quantities
 \begin{equation}
 \rho_s = \frac{m\,f}{M_P}\,R_s, \qquad 
 \alpha = \frac{m\,f}{M_P}\,a, \qquad
 \mu_s = \frac{m}{M_P\,f}\,M_s,
 \end{equation}
 in analogy with the radius and mass of an ASt in eq. (\ref{Scaling}). The constant $c_1$ is always $<0$, and very large: for the sun, $R_s=R_\odot = 7\times10^{5}$ km and $M_s = M_\odot = 2\times10^{30}$ kg, which implies $\rho_\odot = 1735$ and $\mu_\odot = 1.6\times10^{12}$, and then $c_1\sim-10^9$ regardless of the value of $\alpha$.

Evaluating the integral in eq. (\ref{EGScalc}), we have
\begin{align} \label{GSeq}
 E_{GS}(R) &= \frac{3\,G\,M_s\,N\,m}{2\,R_s}\Big(
	  \frac{2\pi^4-50\pi^2+315}{5\pi^2(2\pi^2-15)}\frac{R^2}{R_s^2}\Big) + C_1 \nn \\
	  &= m\,N\,\delta\Big[\frac{\mu_s\,F}{\rho_s^3} \rho^2 + c_1\Big]
\end{align}
where 
\[
F = \frac{3(2\pi^4-50\pi^2+315)}{10\pi^2(2\pi^2-15)} \approx .10.
\]  
 Combining eqs. (\ref{Energyeq}) and (\ref{GSeq}), we can write the rescaled total energy
\begin{equation} \label{eGSTot}
 E_s(\rho) = m\,N\,\delta\,\Big[\frac{A}{\rho^2}
	-\frac{B\,n}{\rho}
	-\frac{C\,n}{\rho^3}
	+\frac{D\,\delta\,n^2}{\rho^6}
	+ \frac{F\,\mu_s}{\rho_s^3} \rho^2 
	+ c_1 \Big].
\end{equation}
The constant $c_1$ affects only the magnitude of $E_s(\rho)$, and not the existence of a minimum, i.e. it will have no effect on our analysis of collapse in the sections below.

\subsection{Collision Rates} \label{CollRatesStarSec}
 When an ASt collides with an ordinary star, $i$ in eq. (\ref{rate}) refers to stellar matter. We will take the number of stars in the Milky Way to be $N_\odot=10^{11}$. Taking the sun as a ``typical'' star, we use the radius $R_\odot = 7\times10^{5} \text{ km}$ in the cross section; we assume for now that stars also move at a virial velocity $\bar{v} \approx 300$ km/s. The corresponding enhancement factor in the cross section is 
 \[
  \frac{2\,G\,(M_\odot+M_{AS})}{\bar{v}^2\,(R_\odot+R_{AS})} \approx 4.25.
 \]
 Plugging everything in, the constant density estimate of the collision rate using eq. (\ref{rate}) is 
 \begin{equation} \label{StarAppx}
  \hat{\Gamma}_{\odot} \approx 400\, \frac{\Fcal_{DM}}{\Fcal_{AS}} 
	    \frac{\text{collisions}}{\text{year}\cdot\text{galaxy}}.
 \end{equation}
 
 As before, we can improve on our estimation of the rate by allowing ASts to be distributed according to the NFW profile in eq. (\ref{NFW}). For the distribution of stars, we use the phenomenological fit found in \cite{MilkyWay} assuming cylindrical symmetry; the authors used a double exponential to describe the disk and a power law for the baryonic halo component. Here we neglect the halo component, which constitutes a few percent correction and is thus negligible at the level of precision of this work. The distribution is
 \begin{align}
   n_{\odot}(\ell,z) &= n_d(L_\odot,0)\exp\Big(\frac{L_\odot}{L_1}\Big)
	      \exp\Big(-\frac{\ell}{L_1}-\frac{z+Z_\odot}{H_1}\Big)\nn \\
	      &+ f\,n_d(L_\odot,0)\exp\Big(\frac{L_\odot}{L_2}\Big)
	      \exp\Big(-\frac{\ell}{L_2}-\frac{z+Z_\odot}{H_2}\Big).
 \end{align}
 In this section we work in cylindrical coordinates, with $\ell=\sqrt{x^2+y^2}$ the distance from the galactic center in the galactic plane, and $z$ the height above the galactic plane. These expressions depend on the fit parameters $L_{1,2}$, $H_{1,2}$, and $f$, as well as the position of the sun $(L_\odot,Z_\odot)$. We use the best fit values reported in \cite{MilkyWay} and ignore uncertainties; these values are reproduced in Table \ref{MWtab}. The sample in this reference contained only about $10^8$ stars, whereas the total number in the Milky Way is understood to be closer to $10^{11}$. To remedy this, we rescale the final parameter in the fit, $n_d(L_\odot,0)$, by requiring the total number of stars to be
 \[
  \int n_\odot\,d^3r = N_\odot = 10^{11},
 \]
 which implies 
 \[
  n_d(L_\odot,0) \approx .15 \text{pc}^{-3}.
 \] 

 The collision rate of ASts with ordinary stars using eq. (\ref{GammaGeneric}) is then\footnote{We assume also a $z\leftrightarrow -z$ symmetry, allowing us to integrate only on $z\in\{0,R_{vir}\}$ and multiply by $2$.}
 \begin{equation}
  \Gamma_\odot = 2\int_0^{R_{vir}}\,dz\,\int_0^{R_{vir}}\,d\ell \,2\pi\,\ell\,
	  \Big(\frac{\Fcal_{DM}\rho_{NFW}(\sqrt{\ell^2+z^2})}{\Fcal_{AS} M_{AS}}\Big)
	  n_\odot(\ell,z) \langle\sigma\,v\rangle_\odot.
 \end{equation}
 For simplicity, we will not take into account the velocity dispersion of stars and use only the virial velocity $v_{vir}$. 
 Then is it easy to compute
 \[
  \langle\sigma\,v\rangle_\odot \approx \sigma_\odot \, v_{vir}
	  = 2.4\times10^{15}\eta \, \frac{\text{km}^3}{\text{sec}}.
 \]
 Performing the integrals over $\ell$ and $z$ gives the result\footnote{In this expression, we have neglected the uncertainty of $R_{vir}$, represented by additional factors of $\eta$, in the integration bounds. This approximation would not be appropriate if $\eta \gg 1$.}
 \begin{equation} \label{RateSun}
  \Gamma_\odot = 3000 \, \eta\,\frac{\Fcal_{DM}}{\Fcal_{AS}}
	    \frac{\text{collisions}}{\text{year}\cdot\text{galaxy}}.
 \end{equation}
 Comparing this result to the approximation in eq. (\ref{StarAppx}), we see that in this case, taking the ASt and stellar distributions into account increases the rate by nearly an order of magnitude. This likely reflects the fact that the distibution of stars in the Milky Way is very far from a constant density; stars are packed very densely into the stellar disk, and the density is peaked strongly at low $z$. Collisions occur with a much higher frequency near the galactic center, where the density of both ASts and ordinary stars is very large; this fact is not captured by the constant-density approximation.
 
 Finally, note that the estimate in eq. (\ref{RateSun}) is likely an underestimate of the total collision rate of ASts with ordinary stars. This is because in the estimate of the cross section, we took the sun as a ``typical'' star, which is likely a reasonable approximation for the $\Ocal(10^{11})$ stars contained in our sample. However, many stars have radii $R_s$ much larger than $R_\odot$ of the sun, and the cross sections for those stars is enhanced by a factor of roughly $(R_s/R_\odot)^2\gg 1$. We have also ignored smaller stars, like red dwarfs, but their contribution to the collision rate is likely to be a small correction.

 
 

 \subsection{Collapse}
 If an ASt collides with an ordinary star, the critical particle number changes as a result of the additional gravitational interaction in eq. (\ref{eGSTot}). Recall that outside the star, there exists a dilute state only for $n<n_c\approx 12.6$. This effective critical particle number decreases in the star. For the sun, we find that the critical particle number decreases to $n_{c,\odot}\approx11.29$, which implies that ASts with $n_{c,\odot}\leq n<n_c$ will collapse when they enter the sun. For $n<n_{c,\odot}$, the dilute state binding energy increases as the ASt moves into the star (though it is still appropriate to consider it a weakly bound state).
 
 \begin{figure}[t]
  \begin{center}
   \includegraphics[scale=.75]{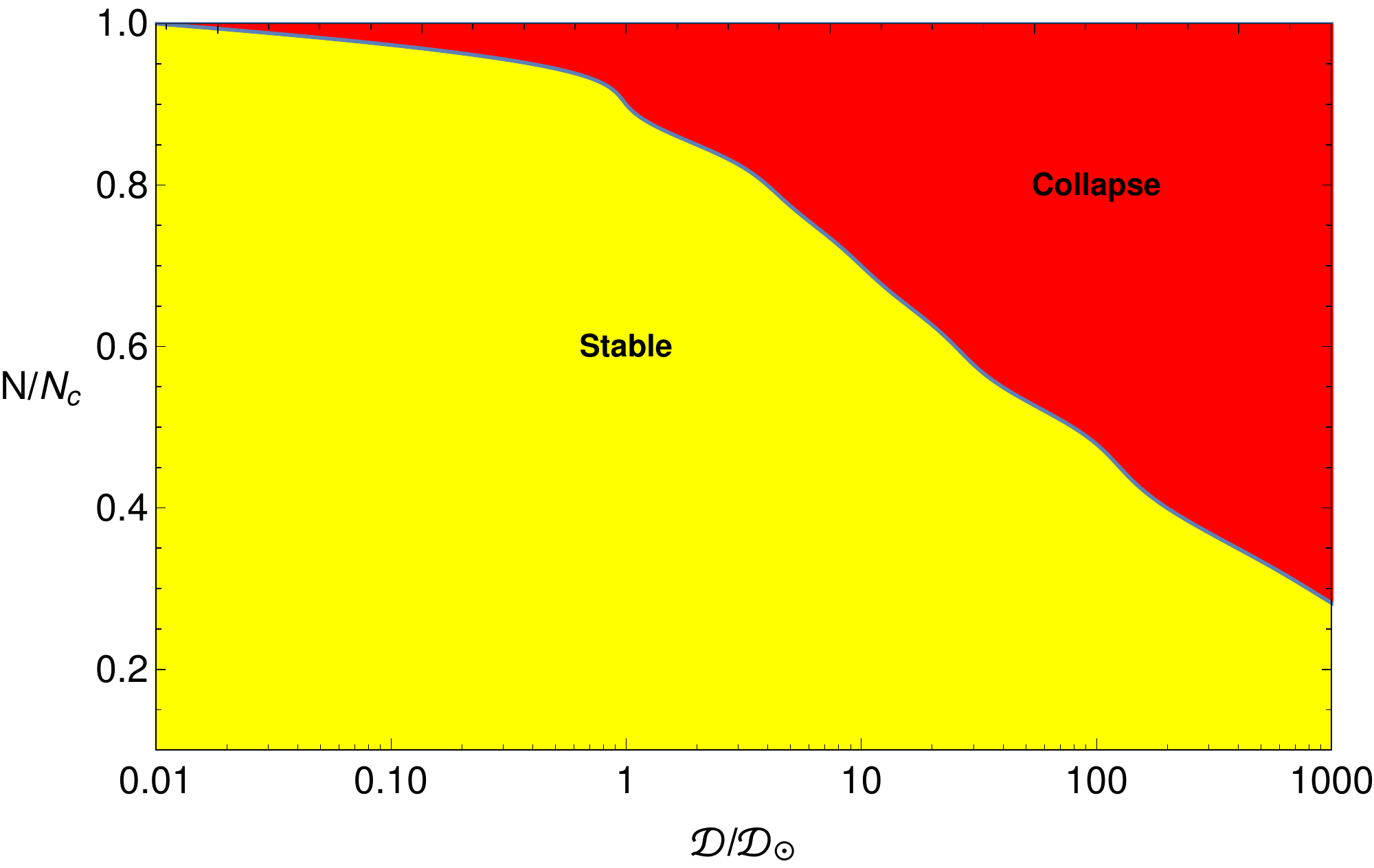}
  \end{center}
   \caption{The parameter space for ASt collapses induced by collisions with ordinary stars. We identify two regions parameterized by the ratio of particle number to critical ($N/N_c$), and density compared to stellar density ($\Dcal/\Dcal_\odot$): A stable region with bound, dilute energy minima (yellow), and a collapse region with no dilute minimum (red).}
   \label{AS_Star_Regions}
  \end{figure}
 
 More generally, the energy landscape in eq. (\ref{eGSTot}) is a function of ASt particle number $N$, and a density parameter $\Dcal\equiv M_s/R_s^3$ characterizing the star. In Figure \ref{AS_Star_Regions}, there exists a region in which ASts colliding with ordinary stars remain stable, though with larger binding energy (shown in yellow), and one in which they collapse (red). In the red region, collapse proceeds via the mechanism outlined in \cite{ELSW}: an initial slow roll, followed by a quick final collapse towards a strongly bound dense state. As the binding energy increases rapidly in the last moments, a large number of relativistic axions are emitted.
 
 If the star has a sufficiently large radius, or if the relative velocity is sufficiently low, then collisions occurring in the red region of Figure \ref{AS_Star_Regions} allow the ASt enough time to collapse fully before it passes fully through the star. To test whether this is plausible, we have used eq. (\ref{tColl}) to calculate the time $t_c$ needed for the ASt to fully collapse. For axion stars with $N = .9\, N_c$ colliding with sun-like stars, this time is $t_c = 47$ mins. 
 
 We compare this to the time a colliding ASt spends inside the star, $t_{in}$. Averaging over the impact parameter of the collision, the average distance an ASt travels through a star is $\bar{d} = 4R_s/3$; the transit time of an ASt moving at a typical velocity $\bar{v}=300$ km/s is roughly
 \[
  t_{in} = \frac{\bar{d}}{\bar{v}} \approx \frac{4\,R_\odot/3}{\bar{v}} = 52 \text{ min},
 \]
 so $t_c<t_{in}$. Thus, on the assumption that the sun is a ``typical'' star, and that a ``typical'' ASt has $N=.9\,N_c$, nearly all collisions represented by the rate in eq. (\ref{RateSun}) result in ASt collapse. This implies a collapse rate of\footnote{Because this conclusion depends on the speed of ASts, we set $\eta=1$ in this expression.}
 \begin{equation}
  \Gamma_{collapse} \sim 3000 \,\frac{\Fcal_{DM}}{\Fcal_{AS}}
	    \frac{\text{collapses}}{\text{year}\cdot\text{galaxy}}.
 \end{equation}
 
 As we pointed out in Section \ref{CollRatesStarSec}, eq. (\ref{RateSun}) is likely an underestimate of the true collision rate. For the same reason, we feel our conclusion that ASts colliding with ordinary stars will collapse is robust, because a larger stellar radius implies a longer transit time, and thus a greater time allowed for collapse. Finally, ASts which do not collapse fully on a single pass through a star also can dissipate kinetic energy through interactions with stellar matter, becoming bound to the star, resulting in repeated collisions and an increased probability for collapse.
 
  
\section{Collisions of an ASt with a Neutron Star} \label{NSsec}
Axion star collisions with neutron stars are particularly interesting, because the strong magnetic fields in the latter can induce currents in the former, leading to stimulated axion conversion to radio-frequency photon bursts that could be observed. This mechanism has been proposed as the source of observed Fast Radio Bursts (FRBs) \cite{Lorimer,Keane,Thornton,Spitler}, and recent estimates suggest that the rate of collisions of ASts with neutron stars is compatible with the observed frequency of FRBs \cite{TkachevFRB,IwazakiFRB,Raby}. In this section we repeat the analysis of previous authors using eq. \eqref{rate}, but also improve on these estimates by taking into account the NFW distribution of ASts, and a nontrivial distribution of neutron stars as well.

If ASts and neutron stars were distributed with uniform density in the galaxy, the rate of collisions is that of eq. \eqref{rate}. The gravitational enhancement of the cross section is
\[
 \frac{2\,G\,(M_{NS}+M_{AS})}{\bar{v}^2\,(R_{NS}+R_{AS})} \approx 2\times10^{4},
\]
where we used $R_{NS} = 10$ km and $M_{NS} = 1.4 M_\odot$. This is an enormous enhancement reflecting the fact that the neutron star is very compact and massive. The resulting cross section is
\[
 \sigma_{NS} = 2.7\times10^9 \text{ km}^2.
\]
Using the value $N_{NS} = 10^9$ for the number of neutron stars, we find a collision rate of
\begin{equation} \label{NSappx}
 \hat{\Gamma}_{NS} = 1.5\times10^{-3} \frac{\Fcal_{DM}}{\Fcal_{AS}} 
	      \frac{\text{collisions}}{\text{year}\cdot\text{galaxy}}.
\end{equation}
This result was already known by previous authors \cite{TkachevFRB,IwazakiFRB,Raby} to be of the right order of magnitude to account for the reported frequency of observed FRBs.

As with other astrophysical collisions, we can improve on this estimate by taking nontrivial distributions into account. For ASts, we use (again) the NFW profile in eq. \eqref{NFW}, while for neutron stars we use the phenomenological fit of \cite{NSdist}. The number density $n_{NS}(\ell,z)$ of neutron stars can be written in terms of two probability distributions $n_{NS}(\ell,z) = \Ncal\,p_\ell(\ell)\,p_z(\ell,z)/2\pi \ell$, where
\begin{align}
 p_\ell(\ell) &= A_{0,\ell} + A\frac{\ell^{\alpha-1}}{\lambda^\a} 
	e^{-\ell/\lambda} \text{ and} \nn \\
 p_z(\ell,z) &= A_{0,z}\, \Theta(z - .1\text{ kpc}) + A_1\, e^{-z/h_1(\ell)} 
	+ A_2\, e^{-z/h_2(\ell)},
\end{align}
where $\Theta(x)$ is a Heaviside function. The scale heights $h_{1,2}(\ell)$ are defined by
\begin{align}
 h_1(\ell) &= k_1\,\ell + b_1 \nn \\
 h_2(\ell) &=   \begin{dcases}
    k_2^<\,\ell + b_2^< ,
	      & \ell\leq 4.5 \text{ kpc}\\
    k_2^>\,\ell + b_2^> ,
	      & \ell\geq 4.5 \text{ kpc}
  \end{dcases}
\end{align}
The coordinates $\ell$ and $z$ are defined as in Section \ref{CollRatesStarSec}, and the remaining constants are best-fit parameters from the analysis of \cite{NSdist}, reproduced in Table \ref{MWtab}. We normalize the distribution using $\Ncal$ by the requirement
\[
 2 \int_0^{R_{vir}}\int_0^{2\pi} \int_0^{R_{vir}}\,\ell\,n_{NS}(\ell,z) 
	  \,d\ell\,d\theta\,dz = N_{NS} = 10^9.
\]

We again use $v_{vir} = \eta\,\bar{v}$, and ignore velocity dispersion so that $\langle\sigma\,v\rangle\approx\sigma\,v_{vir}$. We find the total collision rate to be
\begin{align} \label{NSfull}
 \Gamma_{NS} &= 2\,\Ncal\,\int_0^{R_{vir}}\,dz\,\int_0^{R_{vir}}\,d\ell 
	  \Big(\frac{\Fcal_{DM}\rho_{NFW}
	  (\sqrt{\ell^2+z^2})}{\Fcal_{AS} M_{AS}}\Big)
	  p_\ell(\ell)\,p_z(\ell,z) \langle\sigma\,v\rangle_{NS} \nn \\
	  &= \frac{2\,\Ncal\,\sigma_{NS}\,v_{vir}}{M_{AS}}\frac{\Fcal_{DM}}{\Fcal_{AS}}
	      \int_0^{R_{vir}}\,dz\,\int_0^{R_{vir}}\,d\ell \,
	      \rho_{NFW}(\sqrt{\ell^2+z^2})\,p_\ell(\ell)\,p_z(\ell,z) \nn \\
	  &= 4\times10^{-4} \, \eta\,\frac{\Fcal_{DM}}{\Fcal_{AS}}
	    \frac{\text{collisions}}{\text{year}\cdot\text{galaxy}}.
\end{align}
It should be noted that an estimate of this kind, using the neutron star distribution found in \cite{NSdist}, was first performed in \cite{BarrancoNS} assuming spherical symmetry, whereas we have included the $z$-dependence in $n_{NS}$.

In this work we have remained agnostic about how axion stars form, preferring to parameterize our ignorance by the parameters $\Fcal_{AS}$ and $\Fcal_{DM}$. It is sometimes argued that all axion stars are formed in the early universe through the collapse of axion miniclusters. However, it is also possible that axion stars continually form as the dilute background of axions thermalizes and clumps together; such accretion of axions into condensates might be especially efficient if they form in the cores of ordinary stars. In this analysis, we have strived to correctly capture the relevant physics, as well as the astrophysical uncertainties regarding axion star formation. Because of these important uncertainties, we do not comment here on the interpretation of Fast Radio Bursts as originating in ASt collisions with neutron stars.

\section{Conclusions} \label{ConcSec}

We have calculated the collision rates of axion stars (ASts) with each other, with ordinary stars, and with neutron stars, in dark matter halos similar to that of the Milky Way. In each case, we have taken into account the number density distributions of each class of astrophysical bodies, and by analyzing the energy functional, determined whether the ASt could survive the collision, or if it instead collapses. The rates of these collisions are large, some as frequent as $\Ocal(10^7)$ collisions/year/galaxy, but each can be larger or smaller by a few orders of magnitude, due to the uncertainty in the calculations, which is absorbed into the free parameters $\Fcal_{DM}$, $\Fcal_{AS}$, and $\eta$, defined in Section \ref{CollRatesASSec}.

The collision rate of ASts with ordinary stars is of $\Ocal(10^3)$ collisions/year/galaxy. Because an ordinary star is so large and so massive compared to a typical ASt, the gravitational effect it exerts is important during a collision. In particular, stable ASts approaching ordinary stars develop deeper energy minima as they pass through. ASts which are sufficiently close to their critical mass $M_c$ collapse under the added influence of the star's gravitational field, as seen in Figure \ref{AS_Star_Regions}. Keeping in mind the relevant uncertainties in axion star formation and astrophysical data, it is plausible that a very large number of ASt collapses are induced through collisions with ordinary stars.

On the other hand, collisions between two ASts, even with masses very close to critical, are unlikely to result in collapse; this is because a very low relative velocity is required to allow enough time for the ASts to collapse while occupying the same volume. We find that this condition is satisfied only when $v_{rel} \lesssim 1$ km/s, even for the heaviest stable ASts (see Figure \ref{ASAS_time}). The required velocity decreases as the mass of the ASts in question decreases further from criticality. We conclude that in spite of a large total collision rate, the small number of collapses induced in this way will likely have a negligible effect on the dark matter or ASt mass distribution.

As ASts collapse, they can emit relativistic axions in the form of radiation \cite{ELSW,Tkachev2016,Marsh2016}. This is because the binding energy of a collapsing ASt increases rapidly \cite{Lifetime}. It is possible that all, or a large fraction of, the mass of the ASt would be converted in this way. If this is the case then the total axion star contribution to dark matter in galactic halos could be decreasing with time, as relativistic axions escape from galaxies and galaxy clusters. If this decrease is substantial, then it could constrain the hypothesis that axion stars constitute a large fraction of dark matter. This could also lead to interesting possibilities for astrophysical observations, or other cosmological consequences; we will return to these topics in a future work.

We find for both of the above types of collisions that, over the majority of parameter space, only those axion stars with masses very close to the critical value $M_c$ can collapse through collisions. Because neither the primordial nor current mass distribution of axion stars in dark matter halos is known to a high degree of certainty, it is unclear precisely how many such ASts should be expected. That said, we find a high rate for collapses induced by ASt collisions with ordinary stars; this makes the collapse possibility interesting. Even if only a small fraction of dark matter is in the form of near-critical ASts (i.e. if $\Fcal_{DM} \ll 1$), it is possible that their collapse and subsequent conversion to relativistic axions could lead to observable consequences. This question could be fully answered by a dedicated study of the formation mechanism and timescale of ASts, a task we plan to undertake in the near future.

Using the same framework, we have also revisited previous calculations of the collision rate betwen ASts and neutron stars. Such events are of particular interest because they can result in novel detection signatures. We improve on previous estimates by including the full dependence on the equatorial radius $\ell$ and polar distance $z$ of the neutron star distribution. The oft-quoted result of $\G_{NS}\sim10^{-3}$ collisions/year/galaxy can be recovered when using a nontrivial distribution, but only given specific assumptions about the fraction of dark matter in axion stars. We have attempted here to parameterize the uncertainties in this calculation, in the hopes that these estimates can be improved in the future.

Finally, any condensate of bosons will have a maximum mass above which it collapses. Therefore, this framework can be applied to other condensates of bosons with attractive or repulsive interactions, whether they are star-like or galaxy-sized.

\section*{Acknowledgements}
 We thank P. Argyres, J. Berger, P. Fox, R. Gass, R. Harnik, M. Ma, and C. Vaz for conversations. M.L. thanks the WISE program and Professor U. Ghia for support and encouragement, and the University of Cincinnati and the Department of Physics for a summer research fellowship. J.L. thanks the MUSE program and the Department of Physics at University of Cincinnati for support and financial assistance. The work of J.E. was partially supported by a Mary J. Hanna Fellowship through the Department of Physics at University of Cincinnati, and also by the U.S. Department of Energy, Office of Science, Office of Workforce Development for Teachers and Scientists, Office of Science Graduate Student Research (SCGSR) program. The SCGSR program is administered by the Oak Ridge Institute for Science and Education for the DOE under contract number DE-SC0014664. J.E. also thanks the Fermilab Theory Group for their hospitality.

\end{document}